# Can Technology Life-Cycles Be Indicated by Diversity in Patent Classifications? The Crucial Role of Variety.

Loet Leydesdorff

University of Amsterdam, Amsterdam School of Communication Research (ASCoR),

P.O. Box 15793, 1001 NG Amsterdam, The Netherlands; loet@leydesdorff.net

**Abstract**

In a previous study of patent classifications in nine material technologies for photovoltaic cells, Leydesdorff *et al.* (2015) reported cyclical patterns in the longitudinal development of Rao-Stirling diversity. We suggested that these cyclical patterns can be used to indicate technological life-cycles. Upon decomposition, however, the cycles are exclusively due to increases and decreases in the variety of the classifications, and not to disparity or technological distance, measured as (1 – *cosine*). A single frequency component can accordingly be shown in the periodogram. Furthermore, the cyclical patterns are associated with the numbers of inventors in the respective technologies. Sometimes increased variety leads to a boost in the number of inventors, but in early phases—when the technology is still under construction—it can also be the other way round. Since the development of the cycles thus seems independent of technological distances among the patents, the visualization in terms of patent maps can be considered as addressing an analytically different set of research questions.

**Introduction**

In a previous study of nine material technologies for photovoltaic (PV) cells, Leydesdorff *et al.* (2015) found a cyclic pattern in Rao-Stirling diversity (Rao, 1982; Stirling, 2007) using the cosine for technological proximity (Jaffe, 1986) and relative frequencies among patent classifications as variety. The cyclic patterns could be recognized by an expert in these technologies as a reflection of the development of technological life-cycles. In this communication, I decompose the cyclic pattern in the diversity in terms of variety and disparity, respectively. The patterns will also be related to other parameters such as the number of patents, inventors, and assignees. The conclusion is that the disparity does not play a role in generating the cycles, since they can also and even more precisely be indicated by a sole measure of the variety such as the Herfindahl-Hirschman or Simpson index. Spectral analysis confirms that only a single component (i.e., variety) drives the cyclic development. Furthermore, the cyclic pattern in the classifications is reflected in the number of inventors, but with a potential delay.

**Data**

Recently, the U. S. Patent and Trade Office (USPTO) and the European Patent Office (EPO) abandoned their respective classification systems of patents in favor of the Cooperative Patent Classifications (CPC). CPC builds on the International Patent Classifications (IPC) of the World Intellectual Property Organization (WIPO), by taking the first four digits from IPC version 8. However, CPC enhances the hierarchically organized IPC (v.8) by making it possible to add technology-specific tags such as for "nanotechnology" (Y01) or "technologies for mitigating climate change" (Y02).

The new classifications thus provide us with the possibility to generate sets of patents representing advanced technologies with a level of precision perhaps comparable only to the medical subject headings (MeSH) of PubMed/Medline in the case of publications (Lundberg *et al.*, 2006; Rotolo & Leydesdorff, in print). We downloaded from USPTO, all patents tagged with Y02E 10/54$ for nine material technologies in PV cells on August 20, 2013 (Y02E 10/541), and for the other eight technologies in October and November 2013 (cf. Shibata *et al.* 2010). The nine technologies and the numbers of patents under study are shown in Table 1.

| CPC | Description | USPTO | Download date |
|---|---|---|---|
| Y02E 10/541 | CuInSe2 material PV cells | 419 | August 20, 2013 |
| Y02E 10/542 | Dye sensitized solar cells | 547 | October 23, 2013 |
| Y02E 10/543 | Solar cells from Group II-VI materials | 302 | November 26, 2013 |
| Y02E 10/544 | Solar cells from Group III-V materials | 882 | November 26, 2013 |
| Y02E 10/545 | Microcrystalline silicon PV cells | 148 | November 26, 2013 |
| Y02E 10/546 | Polycrystalline silicon PV cells | 269 | November 26, 2013 |
| Y02E 10/547 | Monocrystalline silicon PV cells | 1236 | November 26, 2013 |
| Y02E 10/548 | Amorphous silicon PV cells | 759 | November 26, 2013 |
| Y02E 10/549 | Organic PV cells | 1468 | November 26, 2013 |

**Table 1**: Nine material technologies for photovoltaic cells distinguished in the Cooperative Patent Classifications (CPC).

The data is indexed by professionals, so one would expect the distinctions between the nine technologies to be fine-grained and precise. Because some patents are tagged in more than a single category, the 6,030 tags (in the third column of Table 1) are based on a smaller number of patents.

**Methods**

Using VOSviewer (Van Eck & Waltman, 2011) for the visualization, Leydesdorff, Kushnir, and Rafols (2014) generated global maps on the basis of cosine-normalized vectors of the 124 IPC classes at the 3-digit level and of the 630 IPC classes at the 4-digit level. These maps can be used to project the IPCs in specific set(s) of patents under study in terms of both relative frequencies (size of the nodes) and distances on the map. The reader is referred to Leydesdorff, Alkemade, Heimeriks, and Hoekstra (2015) for more details and examples of the mapping and overlay techniques. In this study, we use the cosine values between the vectors of the 630 IPC classes at the 4-digit level.[1]

Rao-Stirling diversity combines two of the three aspects of interdisciplinarity distinguished by Rafols & Meyer (2010): variety and disparity. (The third aspect, balance or coherence, was further developed by Rafols *et al.* (2012) for interdisciplinary units and by Leydesdorff & Rafols (2011) for developments at the field level.) Leydesdorff *et al.* (2013) added the value of Rao-Stirling diversity ($\Delta$) routinely to the output as a measure of interdisciplinarity in the case of journal maps. What may be indicated by this same measure in the case of patent maps?

Rao-Stirling diversity is defined as follows (Rao, 1982; Stirling, 2007; cf. Zhang *et al.*, 2014):

$$\Delta = \sum_{ij} p_i p_j d_{ij} \qquad (1)$$

where $d_{ij}$ is a disparity measure between two classes $i$ and $j$—the categories are in this case IPC classes at the 4-digit level—and $p_i$ is the proportion of elements assigned to each class $i$. As the disparity measure, we use $(1 - cosine)$ since the cosine values of the citation relations among the aggregated IPC were used for constructing the base map. Jaffe (1986, at p. 986) proposed taking the cosine between the vectors of classifications as a measure of "technological proximity." In other words, we do not use the distances on the maps themselves, but the cosine values that were initially used for constructing the maps.

---

[1] The file with the 630 * 630 cosine values can be retrieved at http://www.leydesdorff.net/ipcmaps/cos_ipc4.dbf .

**Technology life-cycles**

Figure 1 shows the development of Rao-Stirling diversity using 419 USPTO-patents in the (first) CPC class under study (that is, Y02E10/541) during the period 1975-2012. This figure suggests that the technology was developed in three cycles.

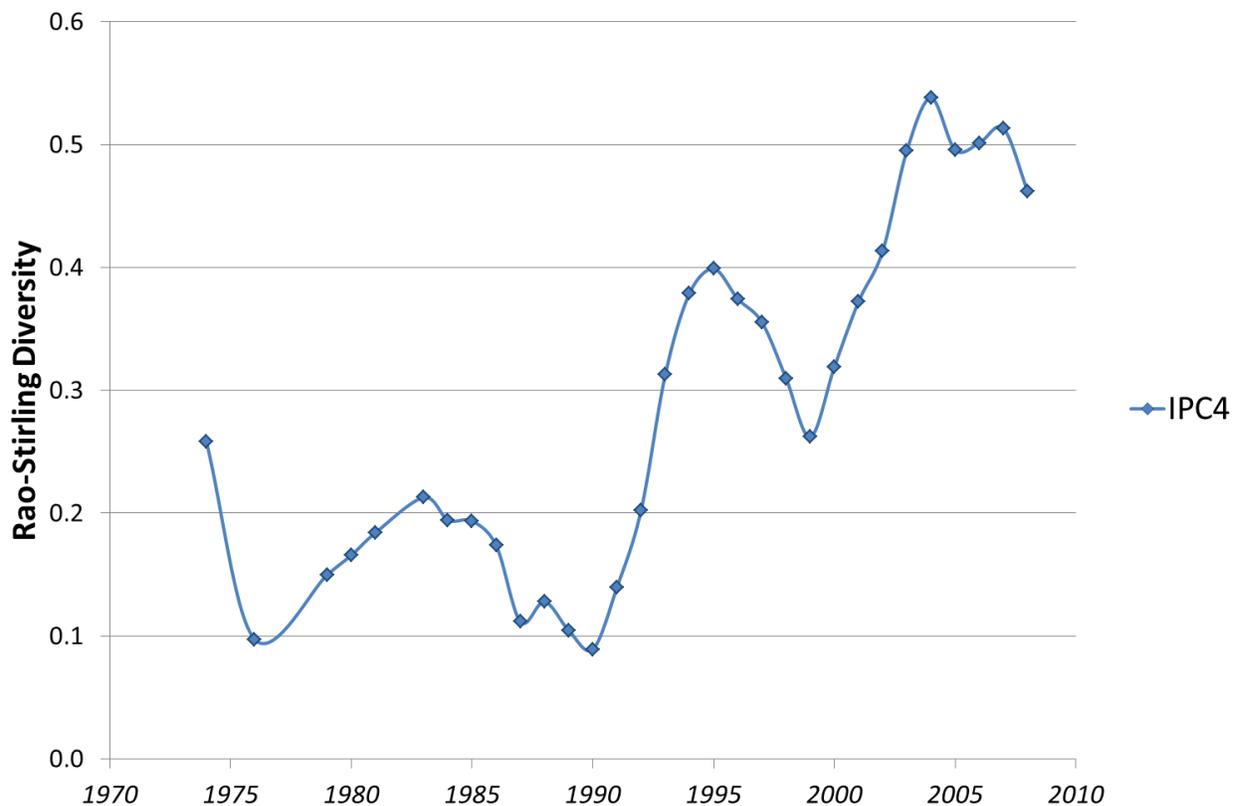

**Figure 1**: The development of Rao-Stirling diversity in IPC (three and four digits) among 419 USPTO-patents with CPC Y02E10/541 ("CuInSe$_2$ material PV cells") during the period 1975-2012.

Two of the valleys, i.e., the period of decreasing diversity in the late 1980s and the most recent such period, correspond with breakthroughs in the efficiency of thin-film solar cells (Green *et al.*, 2013). On the basis of analysis of co-invention addresses, expert interviews, and secondary literature, Leydesdorff *et al.* (2015, p. 640) specified these three cycles as follows (Shafarman & Stolt, 2003):

1. an early cycle during the 1980s which is almost exclusively American; after initial development of the technology at Bell Laboratories in the '70s, Boeing further developed the solar cells using these materials;
2. a second cycle during the 1990s that includes transatlantic collaboration and competition with Europe; the US, however, remains in the lead; and
3. a third and current cycle—the commercial phase—marked by the prevalence of American-Japanese collaboration and by collaboration *within* Europe.

Similar cycles were found using the other eight CPC classes under study.

Since Rao-Stirling diversity is composed of two components (variety and disparity), one can first ask which of the two components carries the cycles; or is it perhaps an interaction? Secondly, the cycles can perhaps be related to other attributes of the respective sets of patents, such as the numbers of patents, inventors, or assignees. Thirdly, one can correlate the longitudinal development of the nine technologies, and ask whether the developments have a single pattern in common; perhaps caused (for example) by changes in the policy of the patent office?

**The decomposition of Rao-Stirling diversity**

If all disparity is equal to one ($d_{ij} = 1$), $\Delta = \sum_{i \neq j} p_i p_j$. This is also called the Gini-Simpson index of diversity, and for analytical reasons, it is the complement to one of the Herfindahl-Hirsch index or equivalently the Simpson index (Stirling, 2007).[2] Figure 2 shows that the variety term under this assumption of all $d_{ij} = 1$ accounts for the cyclic development in Figure 1.

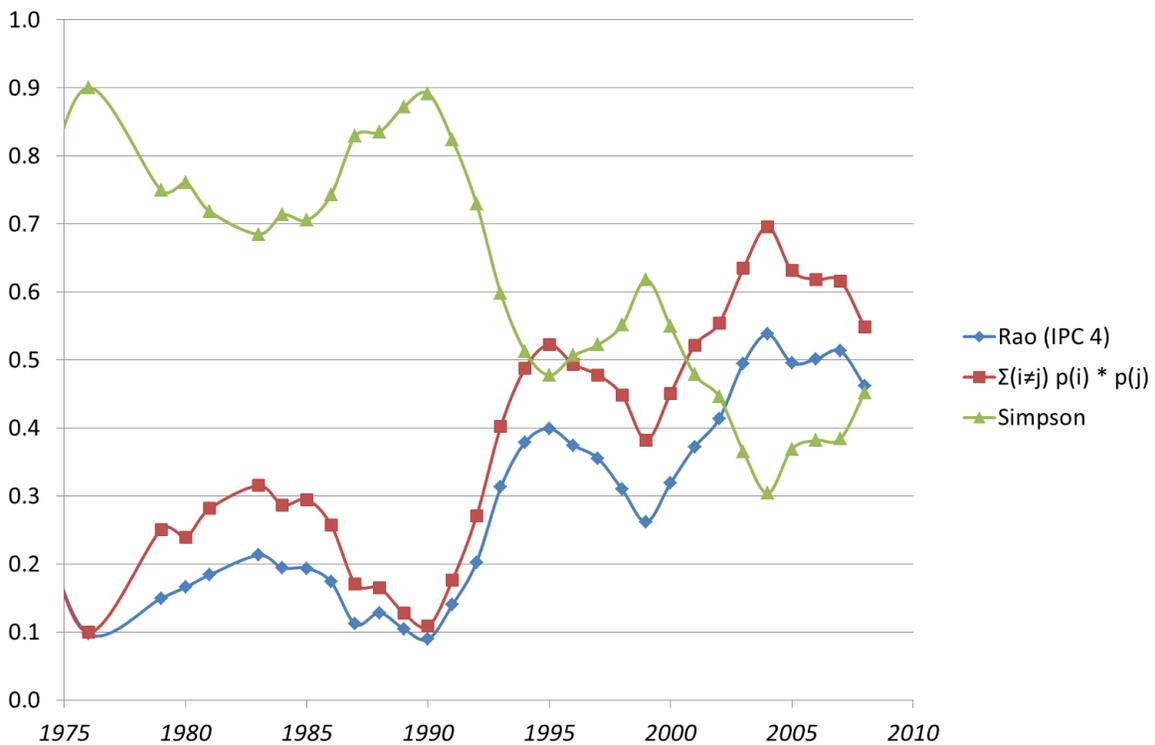

---

[2] $\sum_{ij} p_i p_j = 1$ when taken over all $i$ and $j$. The Simpson index is equal to $\Sigma_i (p_1)^2$, and the Gini-Simpson to $[1 - \Sigma_i (p_1)^2]$.

Furthermore (Zhou *et al.*, 2012, pp. 804f.):

$$\sum_{ij} p_i p_j = \sum_i p_i p_i + \sum_{i \neq j} p_i p_j$$
$$1 = \sum_i p_i p_i + \sum_{i \neq j} p_i p_j$$
$$\sum_i p_i p_i = 1 - \sum_{i \neq j} p_i p_j$$

Or, in other words: Simpson = 1 – variety.

Note that for $i = j$ —that is the diagonal—cosine($i,i$) = 1, and the disparity (1 – $cos$) = 0. Therefore, this term does not contribute to the Rao-Stirling diversity in our case, and variety is equal to $\sum_{i \neq j} p_i p_j$.

**Figure 2**: Rao-Stirling diversity, variety, and the Simpson Index for IPC 4-digit classes in 419 USPTO patents tagged CPC Y02E10/541 ("CuInSe$_2$ material PV cells") during the period 1975-2012.

Figure 2 shows that the cyclic pattern in Rao-Stirling diversity is caused by changes in the variety; the disparity is not needed for the explanation. Multiplication by a disparity measure (1 - *cosine*) attenuates the pattern exhibited using the Simpson (or Herfindahl) index. In sum, the latter indicator can be used for this analysis of diversity. Analysis of variety in the case of the other eight technologies led to similar results.

**Spectral analysis (Periodogram)**

The question of whether one or two components are involved in the cycles can also be addressed using spectral analysis. In order to test this question, I performed spectral analysis of the curve in Figure 1 using SPSS v.22. (Since spectral analysis requires an even number of observations, the first observation (1975) is not used.) Spectral analysis allows for testing an estimated spectrum in descriptive data without any *a priori* constraints (SPSS, 1999, p. 205).

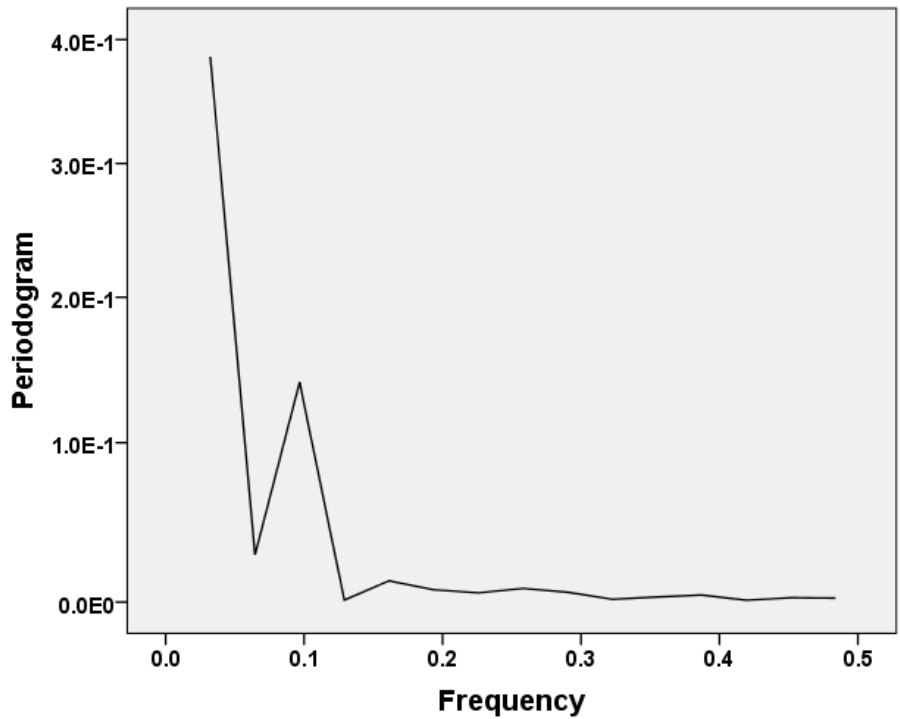

**Figure 3**: Periodogram of the development of Rao-Stirling diversity in IPC (three and four digits) among 419 USPTO-patents with CPC Y02E10/541 ("CuInSe$_2$ material PV cells") during the period 1976-2012. (SPSS v.22).

The remaining 30 observations exhibit a single frequency at 0.1 (Figure 3), indicating that three cycles are involved (3/30 = 0.1). The upshot on the left side of the figure indicates a linear trend—upward as visible in Figure 1. De-trending the curve of Figure 1 (using difference between consecutive years) provides Figure 4.

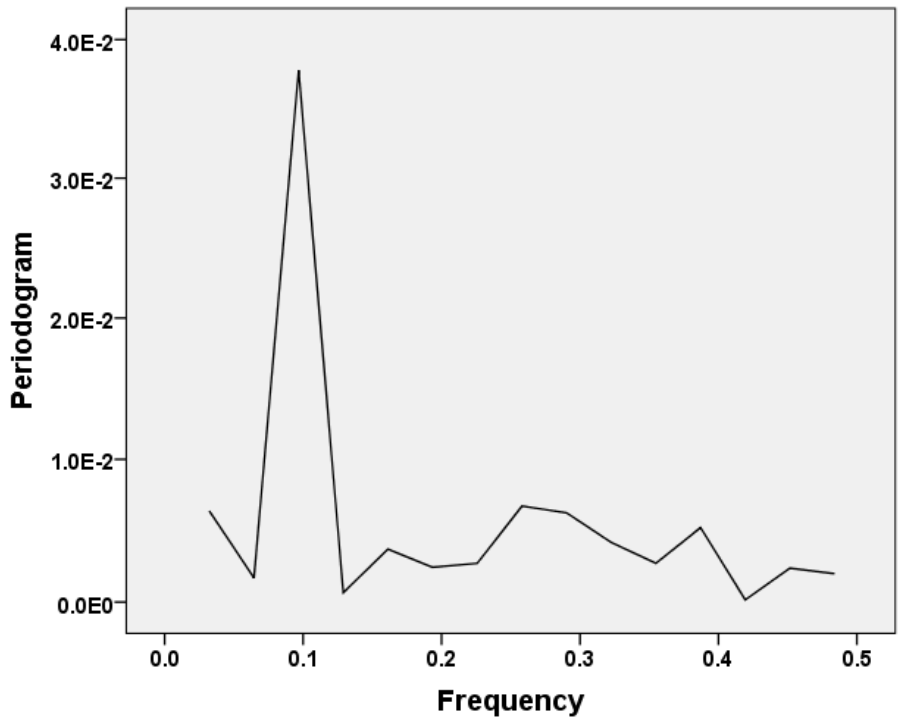

**Figure 4**: De-trended periodogram of the development of Rao-Stirling diversity in IPC (three and four digits) among 419 USPTO-patents with CPC Y02E10/541 ("CuInSe$_2$ material PV cells") during the period 1976-2012. (SPSS v.22.)

This result confirms that a single component drives the cycles. This single component was identified above as variety.

**Other parameters**

Figure 5 shows that the numbers of patents and assignees in this set are highly correlated, and both show exponential growth during the period under study. The number of inventors, however, varies more. Patents in this domain (and in the others) tend to be assigned to a single assignee, whereas the number of co-inventors is less restricted.

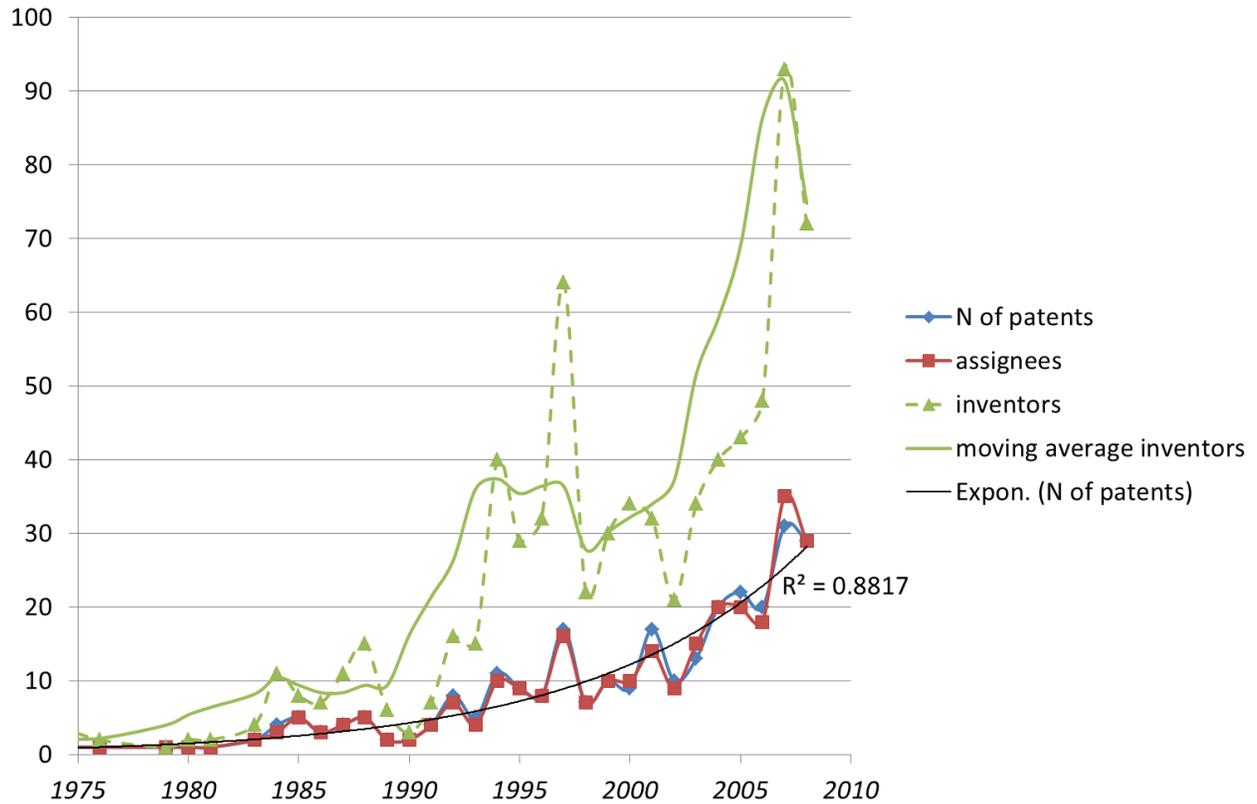

**Figure 5**: Numbers of patents, assignees, and inventors in 419 USPTO patents tagged with CPC Y02E10/541 ("CuInSe$_2$ material PV cells") during the period 1975-2012.

The cyclic pattern in Figure 1 can be retrieved by assuming similarly a five-year moving average (MA) in the number of inventors (Figure 6).

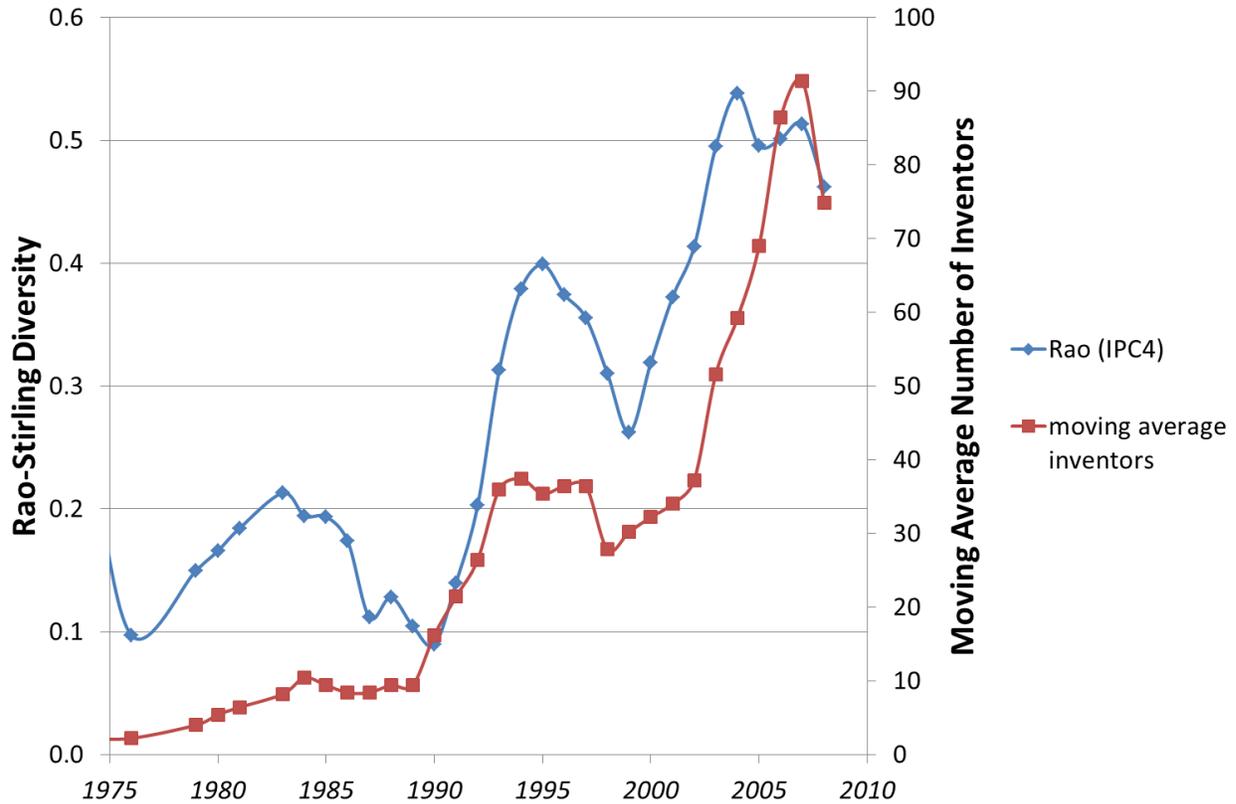

**Figure 6**: Rao-Stirling diversity and the number of inventors for 419 USPTO patents tagged with CPC Y02E10/541 ("CuInSe$_2$ material PV cells") during the period 1975-2012.

Figure 6 shows that the number of inventors lags behind the variety during the last cycle, but not during the valley around 1990. The relative lead of the variety when the volume has grown may indicate that the economic upswing in a technology attracts inventors more than that single inventors are able to induce technological cycles in this more mature stage (Frenken & Leydesdorff, 2000).

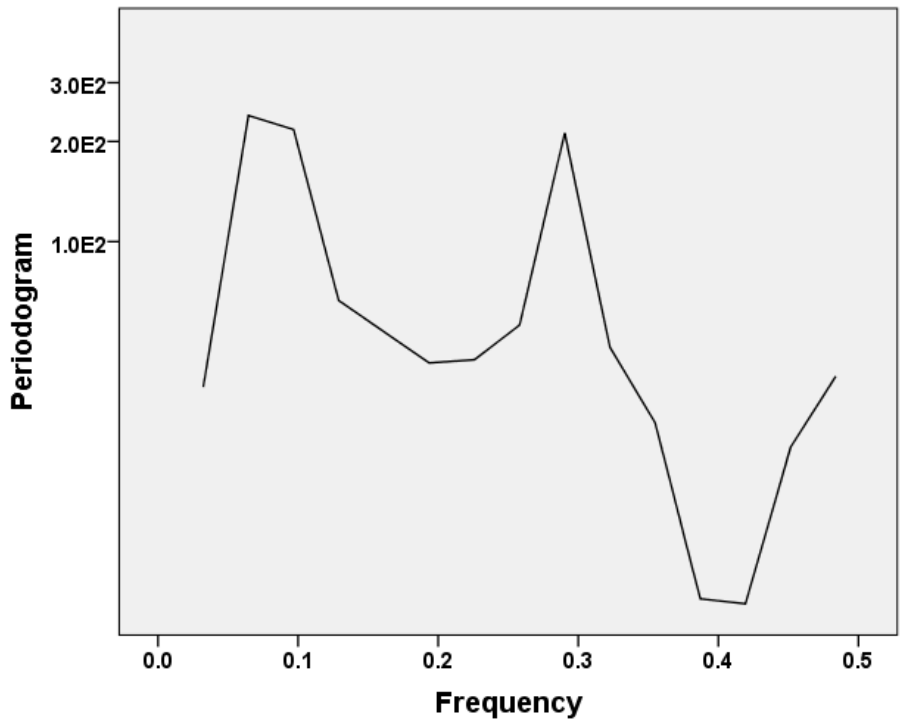

**Figure 7**: De-trended periodogram of the number of inventors for 419 USPTO patents tagged with CPC Y02E10/541 ("CuInSe$_2$ material PV cells") during the period 1975-2012.

The de-trended periodogram of the number of inventors in Figure 7 confirms that a second effect is to be distinguished in this case with a peak at 0.3, and thus indicating nine cycles (9/30 = 0.3). The cycles in the number of inventors can thus be distinguished from longer cycles in the technology.

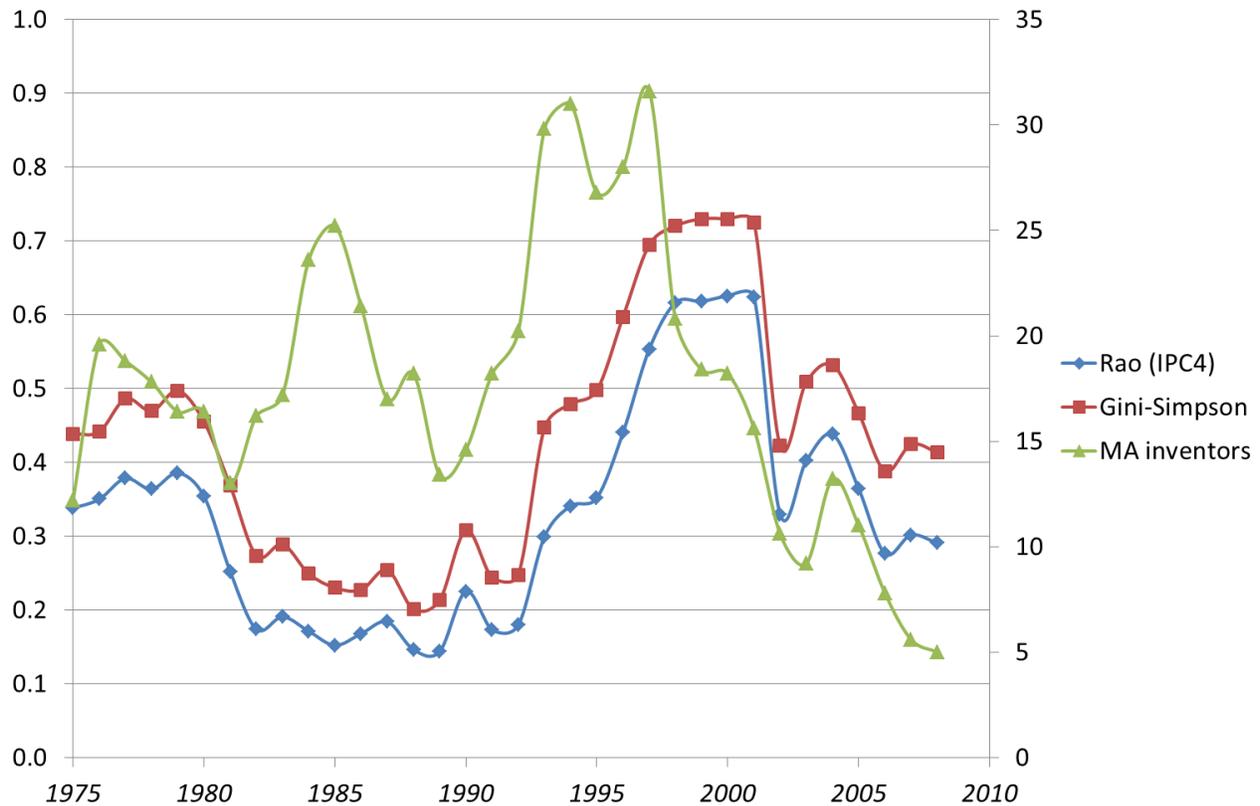

**Figure 8**: Rao-Stirling diversity, Gini-Simpson Index, and (five-year) moving averages of the number of inventors; 320 patents classified Y02E 10/543 ("Solar cells from Group II-VI materials") in USPTO during the period 1975-2012.

In the case of Y02E 10/543 ("Solar cells from Group II-VI materials"), for example, the numbers are smaller, and the moving average of the number of inventors leads the curve of the (Gini-Simpson) variety in this case (Figure 8).

**Correlations**

Spearman's rank-order correlation coefficient ($\rho$) can be used to test the degree to which a monotonic relationship exists between two variables (Sheshkin, 2011, at p. 1366). Since the time-series increases monotonically in terms of sequential years, this measure allows us also to test for increasing or decreasing trends (Bornmann & Leydesdorff, 2013).

**Table 2**: Spearman rank-order correlations of time-series for Gini-Simpson coefficients, 1975-2012.

|      | year    | c541    | c542    | c543    | c544    | c545  | c546    | c547    | c548    | c549    |
|------|---------|---------|---------|---------|---------|-------|---------|---------|---------|---------|
| **year** | 1       | .835**  | .480**  | 0.33    | -0.04   | -0.43 | 0.17    | .403*   | .539**  | 0.19    |
| **c541** | .835**  | 1       | .410*   | .539**  | -0.02   | -0.32 | .531**  | .766**  | .625**  | .532**  |
| **c542** | .480**  | .410*   | 1       | .433*   | .408*   | 0.25  | -0.31   | 0.07    | .653**  | 0.17    |
| **c543** | 0.33    | .539**  | .433*   | 1       | .399*   | 0.21  | 0.21    | .721**  | .617**  | 0.26    |
| **c544** | -0.04   | -0.02   | .408*   | .399*   | 1       | 0.18  | -.518** | 0.17    | 0.32    | 0.19    |
| **c545** | -0.43   | -0.32   | 0.25    | 0.21    | 0.18    | 1     | -0.04   | -0.10   | -0.14   | -0.11   |
| **c546** | 0.17    | .531**  | -0.31   | 0.21    | -.518** | -0.04 | 1       | .549**  | 0.01    | .554**  |
| **c547** | .403*   | .766**  | 0.07    | .721**  | 0.17    | -0.10 | .549**  | 1       | .488**  | 0.31    |
| **c548** | .539**  | .625**  | .653**  | .617**  | 0.32    | -0.14 | 0.01    | .488**  | 1       | 0.16    |
| **c549** | 0.19    | .532**  | 0.17    | 0.26    | 0.19    | -0.11 | .554**  | 0.31    | 0.16    | 1       |

**. Correlation is significant at the 0.01 level (2-tailed).
*. Correlation is significant at the 0.05 level (2-tailed).

Table 2 shows the Spearman correlation coefficients for the years since 1975 and the Gini-Simpson coefficients for the nine PV technologies. A number of these technologies (e.g. Y02E 10/541 and Y02E 10/548) show significantly ($p<0.01$) increasing diversity over time. Y02E 10/544 and Y02E 10/546), however, are negatively correlated among them. Whereas the general pattern is one of increase, the indicator also shows differences among these technologies in terms of the Gini-Simpson index.

**Conclusion**

The cyclical patterns in the Rao-Stirling diversity of nine technologically specific sets of patents were exclusively due to increases and decreases in the variety, and not in the disparity. The variety can, for example, be measured using the Simpson or Herfindahl index. The number of inventors is related to the development of the variety, but possibly with a temporal lag. In early phases of the technology, the development of the variety can be expected to lag, but in later stages the numbers of inventors tend to follow the development of variety in the patent classifications. The nine technologies under study, however, exhibit different patterns: when the technology is under construction the inventors tend to generate the variety, whereas in later stages the number of inventors tends to follow the development of the variety. Accordingly, the curve for the (moving average of the) number of inventors show three times as many cycles (in the periodogram) as the technologies (operationalized as patents). In other words, the technology cycles are relatively long (e.g., ten years).

Whereas inventors follow or participate in constructing a research front, assignees can be considered primarily as economic agents who follow another (economic) logic than the technology cycles. Note that these conclusions are based on a specific set of technologies. Further research should show if variety can be used as a measure of technological development more generally. Our results suggest that the invention process has a dynamic of itself that is longer-termed than the cycling in the average number of inventors (Ivanova & Leydesdorff, 2015). The inventors can then be considered as reflexively participating in retaining wealth from technological developments.